\documentclass{article}
\usepackage{amsmath}

\newtheorem{theorem}{Theorem}
\newtheorem{acknowledgement}[theorem]{Acknowledgement}

\input{tcilatex}

\begin{document}

\title{Entanglement-assisted classical information capacity of the amplitude
damping channel}
\author{Xian-Ting Liang\thanks{%
Email address: xtliang@ustc.edu} \\
Department of Physics and Institute of Modern Physics,\\
Ningbo University, Ningbo, Zhejiang 315211, China}
\maketitle

\begin{abstract}
In this paper, we calculate the entanglement-assisted classical information
capacity of amplitude damping channel and compare it with the particular
mutual information which is considered as the entanglement-assisted
classical information capacity of this channel in Ref. 6. It is shown that
the difference between them is very small. In addition, we point out that
using partial symmetry and concavity of mutual information derived from
dense coding scheme one can simplify the calculation of
entanglement-assisted classical information capacities for
non-unitary-covariant quantum noisy channels.

PACS number(s): 03.67. Hk, 03.65.Ta, 89.70.+c

\end{abstract}

Entanglement-assisted classical information capacity of the quantum channel
describes the maximal rate, i.e. information sent per channel usage, when we
use dense coding scheme instead of simple encoding and decoding to transmit
the data through the channel $\varepsilon $. In the scheme, the sender, say
Alice, and receiver, say Bob, share a two-qubit entangled state prior to the
transmission. At first, Alice encodes information to be transmitted in an
entangled state by operating her holding qubit, and then she sends the qubit
through a quantum channel to Bob, finally Bob jointly measure two qubits
(one is sent from Alice and the other is held by him at the beginning of
this scheme) to decode the information. If no noise to be considered the
scheme called dense coding, and can transmit two bit classical information
by sending one qubit. However, in fact quantum noise is always exist. When
we consider the effect of noise, is this scheme still superior to the
traditional simple encoding and decoding scheme? If yes, then how superior
is the scheme to the traditional one? Up to now, these problems could only
be concretely answered by calculating the entanglement-assisted classical
information capacities for some concrete quantum noisy channels. So
developing the method of calculating the capacity is an interesting topic.
This problem was first investigated by Bennett, Shor, Smolin, and Thapliyal
(BSST) in \cite{Bennett et al01}, where the depolarizing and erasure
channels in $d$ dimensions were studied exactly. In Ref. \cite{Bennett et
al02} the same authors proposed a remarkable simple formula for calculating
the entanglement-assisted classical information capacity in terms of the
maximal mutual information between Alice and Bob, and the capacity of the
amplitude damping channel was also investigated. In this paper, we shall at
first recalculate the entanglement-assisted classical information capacity
of amplitude damping channel. Then we shall compare it with the particular
mutual information which is taken as the entanglement-assisted classical
information capacity of the amplitude damping channel in Ref. 6. In
addition, we shall summarize the method for calculating the
entanglement-assisted classical information capacity of this kind of
non-unitary-covariant channels. Let's review BSST theorem \cite{Bennett et
al02} \cite{Holevo01} first.

In the BSST theorem, the entanglement-assisted classical information
capacity $C\left( \varepsilon \right) $ of a quantum noisy channel $%
\varepsilon :$ $B\left( \mathcal{H}\right) \rightarrow B\left( \mathcal{H}%
\right) $ is given by 
\begin{equation}
C\left( \varepsilon \right) =\sup_{\rho }I\left( \varepsilon ,\rho \right) ,
\label{e1}
\end{equation}%
where%
\begin{equation}
I\left( \varepsilon ,\rho \right) =S\left( \rho \right) +S\left( \varepsilon
\left( \rho \right) \right) -S\left( \varepsilon ,\rho \right) .  \label{e2}
\end{equation}%
Here, $S\left( \tau \right) $ denotes von Neumann entropy, $S\left(
\varepsilon ,\tau \right) $ denotes entropy exchange. Nielsen \emph{et al}.
in \cite{NandCbook} proposed a method for calculating the entropy exchange,
namely, 
\begin{equation}
S\left( \varepsilon ,\rho \right) =S\left( \Omega \right) =-tr\left( \Omega
\log \left( \Omega \right) \right) ,  \label{e3}
\end{equation}%
where $\Omega _{ij}=tr\left( E_{i}\rho E_{j}^{\dagger }\right) $, and $E_{i}$
denote Kraus operators of the channel $\varepsilon $. The proof of this
theorem was first given by \cite{Bennett et al02} and then improved by
Holevo in \cite{Holevo01}. However, the calculation of the
entanglement-assisted classical information capacity may still be a
difficult problem for some quantum noisy channels because in order to
maximize the mutual information $I\left( \varepsilon ,\rho \right) $, we
must choose the state $\rho $\ in Eq.(\ref{e1}) over all of the possible
states. Fortunately, $I\left( \varepsilon ,\rho \right) $ is a concave
function so if only we can prove the in question channel being a unitary
covariant channel the calculation become easy \cite{Liang01}. However, some
quantum noisy channels are not unitary covariant, so we cannot calculate
their capacities by simply replace $\rho $ with the maximally mixed state $%
\mathbf{1}/d$ in Eq.(\ref{e2}), where $\mathbf{1}$ is the unitary matrix, $d$
is the dimension of the channel$.$ However, for the non-unitary covariant
channel the concavity and the partial symmetry of $I\left( \varepsilon ,\rho
\right) $ can still be used in the calculation.

In the quantum communication, the following problems are always encountered.
What are the dynamics of an atom which is spontaneously emitting a photon?
How does a spin system at high temperature approach equilibrium with its
environment? What is the state of a photon in an interferometer or cavity
when it is subject to scattering and attenuation? Each of these processes
has its own unique features, but the general behavior of all of them is well
characterized by a quantum operation known as amplitude damping. For the
qubit systems, the evolvement of the amplitude damping can be modeled by
amplitude damping channel. The amplitude damping channel is a non-unitary
covariant channel. In the following, we shall use the concavity and the
partial symmetry of $I\left( \varepsilon ,\rho \right) $ investigate its
entanglement-assisted classical information capacity. The Kraus operators of
amplitude damping channel are 
\begin{equation}
E_{0}=\left( 
\begin{array}{cc}
1 & 0 \\ 
0 & \sqrt{1-\eta }%
\end{array}%
\right) ,\qquad E_{1}=\left( 
\begin{array}{cc}
0 & \sqrt{\eta } \\ 
0 & 0%
\end{array}%
\right) .  \label{e4}
\end{equation}%
Suppose the initial state $\rho \in \mathcal{H}=C^{2}$ is $\rho =\frac{1}{2}%
\left( I+\vec{w}\cdot \vec{\sigma}\right) $, its eigenvalues are 
\begin{equation}
\lambda _{1,2}=\frac{1}{2}\left( 1\pm \sqrt{w_{1}^{2}+w_{2}^{2}+w_{3}^{2}}%
\right) .  \label{e5}
\end{equation}%
When the initial state pass through the amplitude damping channel, it will
become 
\begin{equation}
\rho ^{\prime }=\varepsilon \left( \rho \right) =\frac{1}{2}\left( 
\begin{array}{cc}
1+w_{3}+\eta \left( 1-w_{3}\right)  & \sqrt{1-\eta }\left(
w_{1}-iw_{2}\right)  \\ 
\sqrt{1-\eta }\left( w_{1}+iw_{2}\right)  & \left( 1-\eta \right) \left(
1-w_{3}\right) 
\end{array}%
\right) .  \label{e6}
\end{equation}%
The eigenvalues of $\rho ^{\prime }$ are%
\begin{equation}
\lambda _{1,2}^{\prime }=\frac{1}{2}\pm \sqrt{\left( 1-\eta \right)
^{2}w_{3}^{2}+\left( 1-\eta \right) \left( 2\eta
w_{3}+w_{1}^{2}+w_{2}^{2}\right) +\eta ^{2}}.  \label{e7}
\end{equation}%
By using the Kraus operators of amplitude damping channel and formula $%
\Omega _{ij}=tr\left( E_{i}\rho E_{j}^{\dagger }\right) ,$ we obtain 
\begin{equation}
\Omega =\frac{1}{2}\left( 
\begin{array}{cc}
2-\eta +\eta w_{3} & \sqrt{\eta }\left( w_{1}-iw_{2}\right)  \\ 
\sqrt{\eta }\left( w_{1}-iw_{2}\right)  & \eta \left( 1-w_{3}\right) 
\end{array}%
\right) .  \label{e8}
\end{equation}%
The eigenvalues of $\Omega $ are%
\begin{equation}
\chi _{1,2}=\frac{1}{2}\pm \sqrt{\left( 1-\eta \right) ^{2}+2\eta \left(
1-\eta \right) w_{3}+\eta w_{1}^{2}+\eta w_{2}^{2}+\eta ^{2}w_{3}^{2}}.
\label{e9}
\end{equation}%
From Eq.(\ref{e2}) we can obtain the mutual information as%
\begin{eqnarray}
I\left( \varepsilon ,\rho \right)  &=&I\left( \eta ,\vec{w}\right)   \notag
\\
&=&-\lambda _{1}\log _{2}\lambda _{1}-\lambda _{2}\log _{2}\lambda
_{2}-\lambda _{1}^{\prime }\log _{2}\lambda _{1}^{\prime }  \notag \\
&&-\lambda _{2}^{\prime }\log _{2}\lambda _{2}^{\prime }+\chi _{1}\log
_{2}\chi _{1}+\chi _{2}\log _{2}\chi _{2}.  \label{e10}
\end{eqnarray}%
On one hand, from above results we have $I\left( \eta ,w_{1}\right) =I\left(
\eta ,-w_{1}\right) ,$ and $I\left( \eta ,w_{2}\right) =I\left( \eta
,-w_{2}\right) $, so we see $I\left( \eta ,\vec{w}\right) $ being
symmetrical on points $p\left( w_{1},w_{2}=0,w_{3}\right) $. On the other
hand, it was proven that $I\left( \eta ,\vec{w}\right) $ is a concave
function \cite{Keyl}, so the maximum of $I\left( \eta ,\vec{w}\right) $ must
be restricted on the points $p\left( w_{1},w_{2}=0,w_{3}\right) ;$ the
maximum of $I\left( \eta ,w\right) $ must be included in $I^{\prime }\left(
\varepsilon ,\vec{w}\right) :=I\left\vert _{w_{1},w_{2}=0}\right. \left(
\eta ,w_{3}\right) ,$ namely, $C\left( \eta \right) \subset I^{\prime
}\left( \eta ,\vec{w}\right) =I\left\vert _{w_{1},w_{2}=0}\right. \left(
\eta ,w_{3}\right) $. Further, we can calculate the capacities by taking a
series of $w_{3}^{\prime }$ in different $\eta .$ These $w_{3}^{\prime }$ in
different $\eta $ can be obtained as follows: first we take $I^{\prime
}\left( \eta ,\vec{w}\right) $ derivative with respect to $w_{3}$ as%
\begin{equation}
\frac{dI^{\prime }\left( \eta ,\vec{w}\right) }{dw_{3}}=0,  \label{e11}
\end{equation}%
then we solve $w_{3}$ from Eq.(\ref{e11}) we can obtain a series of $w_{3}$,
which are $w_{3}^{\prime }.$ The numerical result of $w_{3}^{\prime }$ is
shown in Table 1 and their values as a function of $\eta $ are plotted in
Figure 1. The capacities $C\left( \eta \right) $, mutual information $%
I\left( \eta ,w=0\right) $ and the difference of $C\left( \eta \right) $ and 
$I\left( \eta ,w=0\right) $ are also given in the Table 1. We plot the $%
I(\eta ,w=0)$ and $C\left( \eta \right) $ against $\eta $ in Fig. 2. It is
shown that the difference between $C\left( \eta \right) $ and $I(\eta ,w=0)$
is very small and we cannot distinguish them in the figure. In order to
compare them we plot the difference $C\left( \eta \right) -I(\eta ,w=0)$ in
Fig. 3.%
\begin{eqnarray*}
&& \\
&& \\
&&Fig.1 \\
&& \\
&&
\end{eqnarray*}%
\emph{Fig.1 }$w_{3}^{\prime }$\emph{\ versus }$\eta $ \emph{for the
amplitude damping channel, where }$w_{3}^{\prime }$\emph{\ make the mutual
information }$I(\eta ,w_{3})$\emph{\ be capacities }$C\left( \eta \right) $%
\emph{.}%
\begin{eqnarray*}
&& \\
&& \\
&&Fig.2 \\
&& \\
&&
\end{eqnarray*}%
\emph{Fig.2 Capacity }$C\left( \eta \right) $\emph{\ and mutual information }%
$I(\eta ,w=0)$\emph{\ versus }$\eta $\emph{\ for the amplitude damping
channel}$.$\emph{\ Their difference being very small; we cannot distinguish
them using this figure.}%
\begin{eqnarray*}
&& \\
&& \\
&&Fig.3 \\
&& \\
&&
\end{eqnarray*}%
\emph{Fig.3 Difference of capacity }$C\left( \eta \right) $\emph{\ and
mutual information }$I(\eta ,w=0)$\emph{\ versus }$\eta $\emph{\ for
amplitude damping channel.}%
\begin{align*}
& 
\begin{tabular}{|c|c|c|c|c|}
\hline
$\eta $ & $w_{3}^{\prime }$ & $C(\eta ,w_{3}^{\prime })$ & $I(\eta ,w_{3}=0)$
& $C-I$ \\ \hline
0.04 & .020707505 & 1.857993856 & 1.857404993 & .588863e-3 \\ \hline
0.08 & .029451443 & 1.754220384 & 1.753086250 & .1134134e-2 \\ \hline
0.12 & .034204349 & 1.663598636 & 1.662142602 & .1456034e-2 \\ \hline
0.16 & .036476402 & 1.580849799 & 1.579274705 & .1575094e-2 \\ \hline
0.20 & .036918238 & 1.503488311 & 1.501955000 & .1533311e-2 \\ \hline
0.24 & .035871433 & 1.430055143 & 1.428681156 & .1373987e-2 \\ \hline
0.28 & .033529523 & 1.359582064 & 1.358444378 & .1137686e-2 \\ \hline
0.32 & .030001598 & 1.291370839 & 1.290509150 & .861689e-3 \\ \hline
0.36 & .025341559 & 1.224884751 & 1.224304412 & .580339e-3 \\ \hline
0.40 & .019562439 & 1.159688417 & 1.159362804 & .325613e-3 \\ \hline
0.44 & .012643065 & 1.095410976 & 1.095283308 & .127668e-3 \\ \hline
0.48 & .004530009 & 1.031721423 & 1.031706094 & .15329e-4 \\ \hline
0.52 & -.0048640556 & .9683103674 & .9682939063 & .164611e-4 \\ \hline
0.56 & -.0156655130 & .9048748897 & .9047166920 & .1581977e-3 \\ \hline
0.60 & -.0280492412 & .8411041849 & .8406371958 & .4669891e-3 \\ \hline
0.64 & -.0422541602 & .7766639116 & .7756955885 & .9683231e-3 \\ \hline
0.68 & -.0586084818 & .7111767546 & .7094908497 & .16859049e-2 \\ \hline
0.72 & -.0775716652 & .6441954457 & .6415556220 & .26398237e-2 \\ \hline
0.76 & -.0998074512 & .5751615422 & .5713188441 & .38426981e-2 \\ \hline
0.80 & -.1263199222 & .5033365085 & .4980450000 & .52915085e-2 \\ \hline
0.84 & -.1587322020 & .4276745835 & .4207252951 & .69492884e-2 \\ \hline
0.98 & -.1999403638 & .3465572468 & .3378573979 & .86998489e-2 \\ \hline
0.92 & -.2560072406 & .2571288324 & .2469137502 & .102150822e-1 \\ \hline
0.96 & -.3442467036 & .1530143199 & .1425950071 & .104193128e-1 \\ \hline
\end{tabular}
\\
& \text{Table 1 The }w_{3}^{\prime }\text{, capacities }C\left( \eta \right) 
\text{, mutual information }I\left( \eta ,w=0\right) \text{ and } \\
& \text{the difference of }C\left( \eta \right) \text{ and }I\left( \eta
,w=0\right) \text{ against parameter }\eta \text{ for amplitude } \\
& \text{damping channel.}
\end{align*}

In conclusion, on the one hand, by using the concavity and the partial
symmetry of $I\left( \rho ,\varepsilon \right) $ we investigate the
entanglement-assisted classical information capacity of the amplitude
damping channel. It is shown that the capacities $C\left( \eta \right) $ are
always a little bigger than the mutual information $I(\eta ,w=0),$ which
were taken as the entanglement-assisted classical information capacities of
the amplitude damping channel in Ref. \cite{Liang01}. From the results we
see the difference between $C\left( \eta \right) $ and $I(\eta ,w=0)$,
namely, $C\left( \eta \right) -I(\eta ,w=0),$ is very small for all of the
parameters $\eta .$ Hence, it is convenient and accurate to replace $C\left(
\eta \right) $ with $I(\eta ,w=0).$ On the other hand, we obtained some
insight into the calculation of entanglement-assisted classical information
capacity for non-unitary-covariant channels. We find that the concavity and
some symmetry of $I\left( \rho ,\varepsilon \right) $ for
non-unitary-covariant channels can help one simplify the calculations. In
particular, a unitary covariant channel corresponds to a entirety
symmetrical channel whose entanglement-assisted classical information
capacity can be calculated by simply replacing $\rho $ with the maximally
mixed state $\mathbf{1}/d$ in Eq.(\ref{e2}).

\begin{acknowledgement}
\begin{acknowledgement}
This work was supported by Youth Fund of Ningbo City of China.
\end{acknowledgement}
\end{acknowledgement}

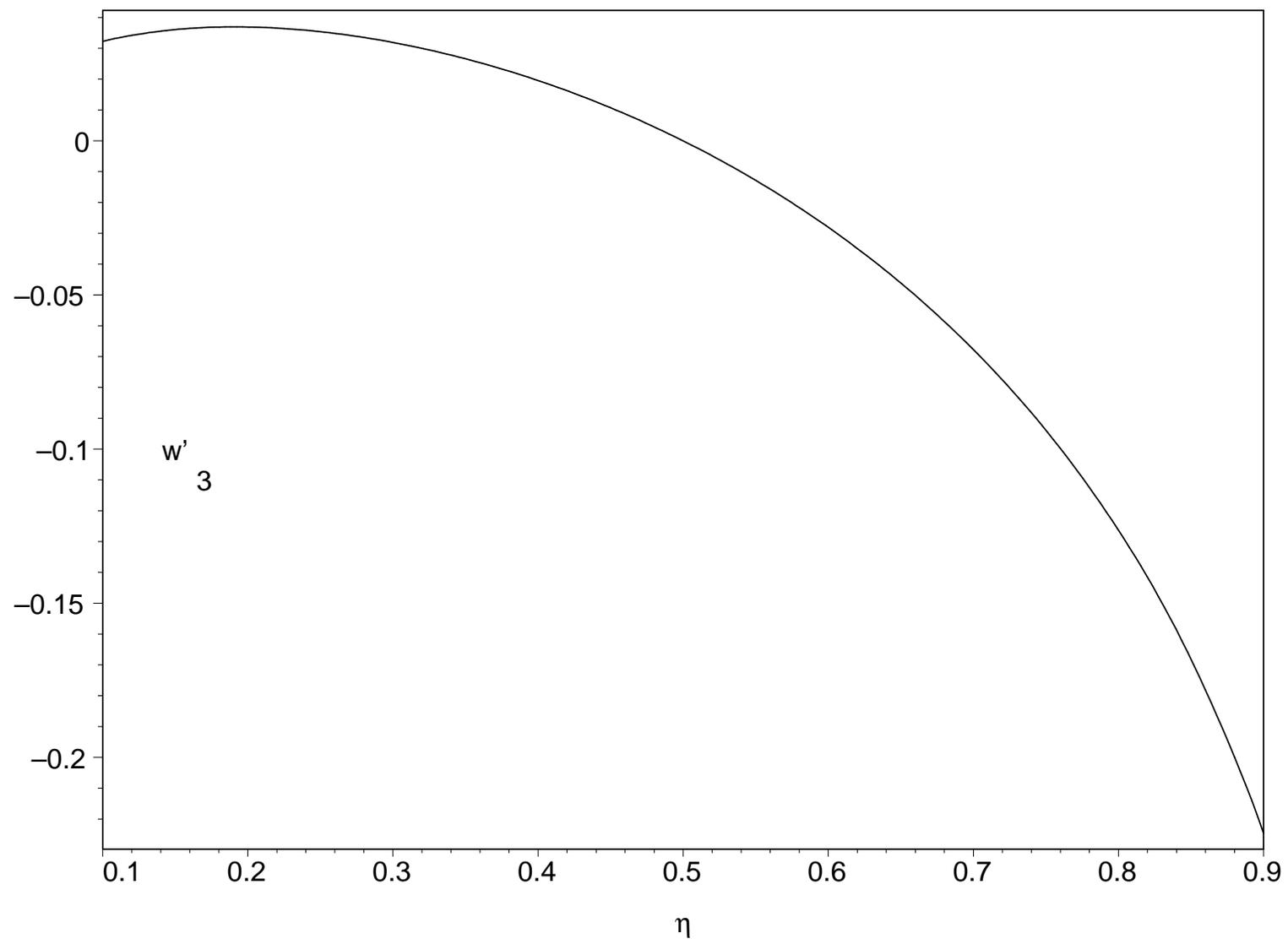
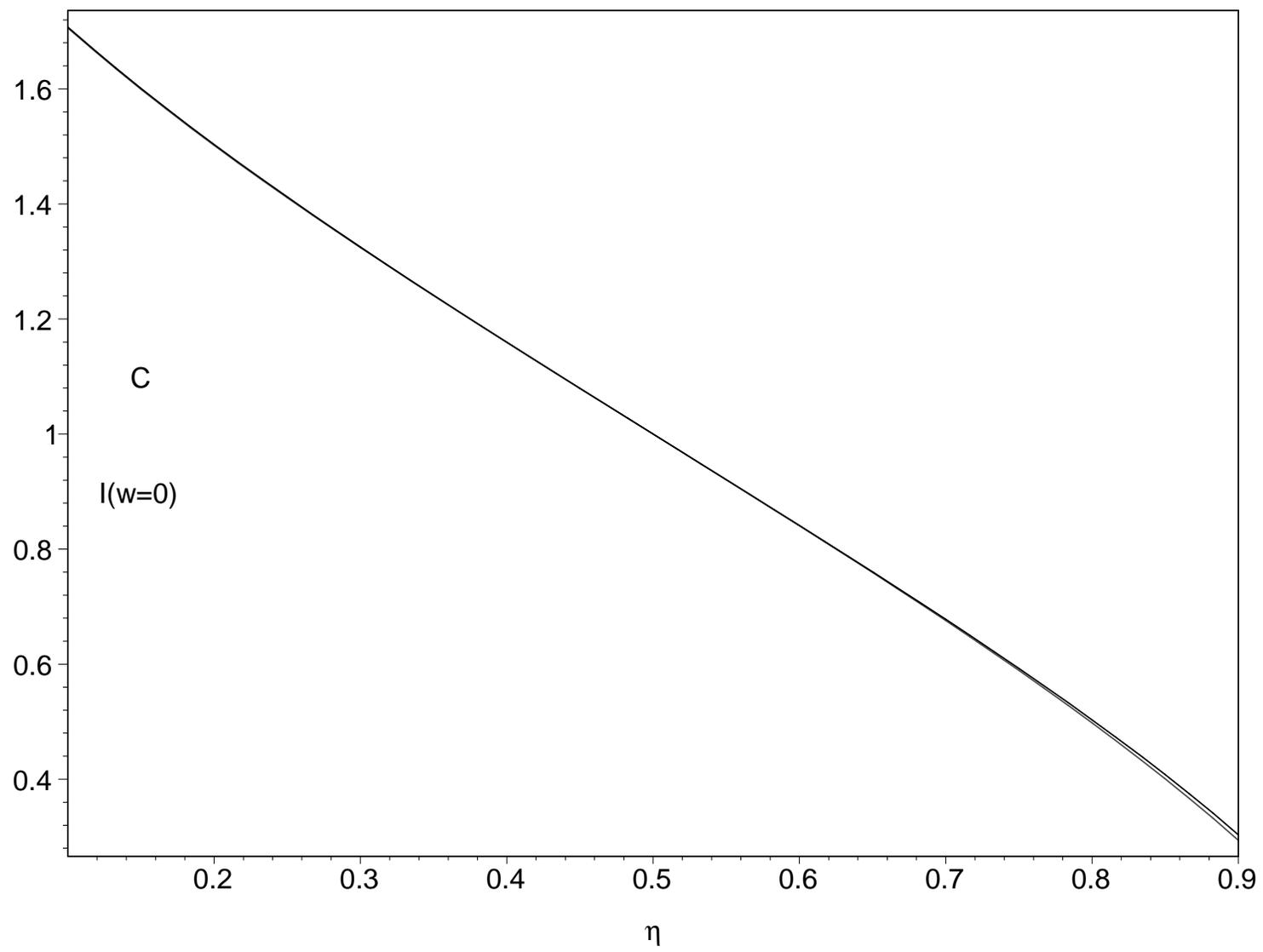
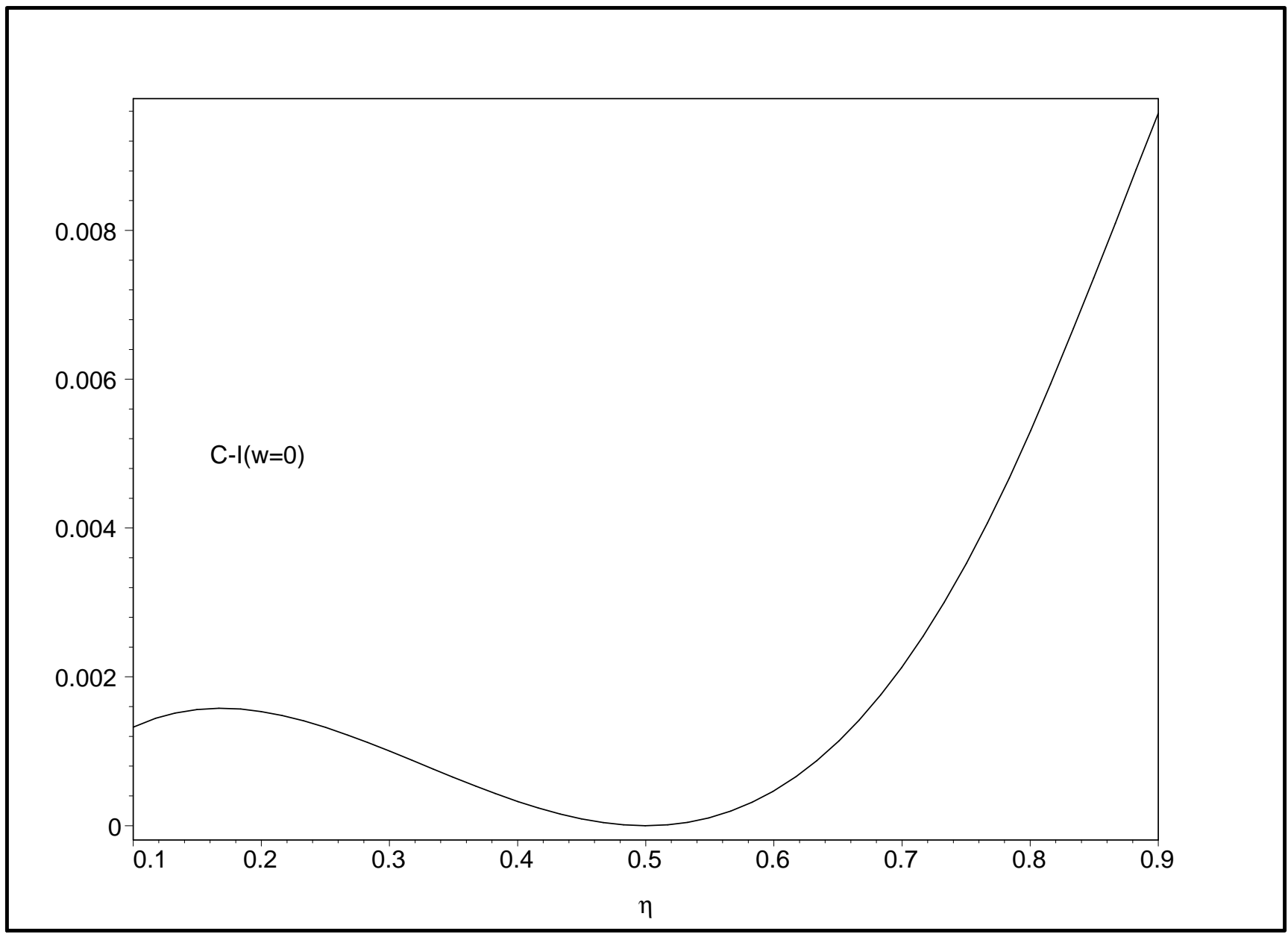
\end{document}